\documentclass[%
reprint,
superscriptaddress,
nofootinbib,
amsmath,amssymb,
prd
]{revtex4-2}

\usepackage{graphicx}
\usepackage{dcolumn}
\usepackage{bm}
\usepackage[dvipsnames]{xcolor}

\begin{document}

\preprint{APS/123-QED}

\title{Muon lifetime and Fermi constant: an update}

\author{Alexander Eberhart}
\affiliation{Institute for Particle Physics Phenomenology, Department of Physics, Durham University, Durham, DH1 3LE, United Kingdom}

\author{Matteo Fael}
\email{matteo.fael@pd.infn.it}
\affiliation{%
Dipartimento di Fisica e Astronomia ``Galileo~Galilei,'' Università di Padova,\\ via F.~Marzolo 8, 35131 Padova, Italy
}%
\affiliation{Istituto Nazionale di Fisica Nucleare, Sezione di Padova, via F.~Marzolo 8, 35131 Padova, Italy
}
\author{Alexander~Keshavarzi}
\affiliation{Department of Physics and Astronomy, University College London, London WC1E 6BT, United Kingdom}
\author{Daisuke~Nomura}
\affiliation{Department of Radiological Sciences, International University of Health and Welfare, Tochigi 324-8501, Japan}
\author{Kay Sch\"onwald}
\email{kay.schonwald@cern.ch}
\affiliation{Theoretical Physics Department, CERN, 1211 Geneva, Switzerland}
\author{Matthias Steinhauser}
\email{matthias.steinhauser@kit.edu}
\affiliation{%
Institut für Theoretische Teilchenphysik, Karlsruhe Institute of Technology (KIT), 76128 Karlsruhe, Germany
}%
\author{Thomas~Teubner}
\affiliation{Department of Mathematical Sciences, University of Liverpool, Liverpool L69 3BX, United Kingdom}
\author{Aidan~Wright}
\email{aidan.wright@liverpool.ac.uk}
\affiliation{Department of Mathematical Sciences, University of Liverpool, Liverpool L69 3BX, United Kingdom}

\date{\today}

\begin{abstract}

We present an updated prediction for the lifetime of the muon
including a detailed analysis of all relevant uncertainties.  Our
prediction includes QED corrections up to order $\alpha^3$ and
state-of-the-art hadronic contributions based on dispersive methods.
Radiative corrections and finite-electron-mass effects are parametrized
by the correction factor $\Delta q$ in
$\tau_\mu^{-1}=G_F^2m_\mu^5(1+\Delta q)/(192\pi^3)$, for which we obtain
$\Delta q=(-4\, 384\, 678 \pm 34)\times 10^{-9}$. This reduces the uncertainty
associated with $\Delta q$ by an order of magnitude compared to the previous
prediction at order $\alpha^2$.
We use our results to provide an updated value of the Fermi coupling constant,
$G_F=1.166\,378\, 59  \, (59) \times 10^{-5} \, \mathrm{GeV}^{-2}$.
Further improvements will require better measurements of the muon lifetime
and mass.

\end{abstract}

\maketitle

\section{Introduction}

The Fermi coupling constant $G_F$ is a fundamental input parameter which
enters the predictions of all quantities involving the electroweak
sector of the Standard Model (SM) of particle physics.  
A numerical value of $G_F$ is obtained from its relation to the muon lifetime $\tau_\mu$ and mass $m_\mu$.
The current experimental precision is dominated by the final MuLan measurement,
$\tau_{\mu^+}=2\, 196\, 980.3 \,(2.2) \, \mathrm{ps}$, which reached a relative uncertainty 
of $1.0$ ppm and implies a determination of $G_F$ at the level of about $0.5$ ppm~\cite{MuLan:2012sih}.

The muon lifetime is conveniently expressed as
\begin{equation}
  \frac{1}{\tau_\mu} =
  \Gamma(\mu^- \to e^- \nu_\mu \bar \nu_e) =
  \frac{G_F^2 m_\mu^5}{192 \pi^3}
  \Big[ 1 + \Delta q \Big]\,.
  \label{eq::tau-GF}
\end{equation}
Higher-order corrections and finite electron-mass effects
are parametrized by the quantity $\Delta q$,
which depends on $\rho= m_e/m_\mu$, the ratio between the 
electron mass, $m_e$, and the muon mass.
By definition, $\Delta q$ contains only corrections to the decay process
$\mu^- \to e^- \nu_\mu \bar{\nu}_e$ evaluated within the Fermi theory. 
Pure electroweak effects associated with the matching of the Fermi theory 
to the SM are absorbed into the definition of $G_F$~\cite{Sirlin:1980nh,Ferroglia:1999tg},
together with tree-level $W$-boson propagator effects~\cite{Ferroglia:2013dga,Fael:2013pja}.
A state-of-the-art evaluation of $\Delta q$ is therefore important for providing 
a robust baseline for the extraction of $G_F$ from the muon lifetime 
and for future high-precision global fits of the electroweak sector.
In this Letter we present an updated result for $\Delta q$, 
including contributions through order $\alpha^3$, and update 
the corresponding prediction for the muon lifetime.

The one-loop QED corrections to $\Delta q$ were computed almost 70
years ago in Ref.~\cite{Kinoshita:1958ru}.  
The two-loop corrections have been
known since the end of the last century from
Refs.~\cite{vanRitbergen:1998yd,Steinhauser:1999bx}.  Results for the
third-order corrections to $\Delta q$ were first  obtained as a
by-product of the computation of the QCD corrections to the
semileptonic $b\to c$ decay~\cite{Fael:2020tow}.  This calculation is
based on an expansion in the mass difference $m_\mu-m_e$ which is
extrapolated to $m_e\to0$, inducing an uncertainty of about
$15\%$ for the coefficient of order $\alpha^3$ of $\Delta q$.  A
dedicated calculation of the fermionic third-order QCD corrections to
$b\to u$ in Ref.~\cite{Fael:2023tcv} can immediately be translated to
muon decay. The combination with the non-fermionic parts from
Ref.~\cite{Fael:2020tow} leads to a reduced uncertainty as will be
described below. In addition, as compared to Ref.~\cite{Fael:2020tow},
more expansion terms in $m_\mu-m_e$ are available.
Results for the QED corrections are presented in Section~\ref{sec::QED}.

At order $\alpha^2$, QCD corrections enter $\Delta q$ via the hadronic vacuum polarization (HVP) of
the photon. Such contributions cannot be computed within
the perturbative framework and non-perturbative methods have to be
applied. 
The current prediction of the hadronic contribution is based on an
analysis performed in 1998~\cite{vanRitbergen:1998hn}.  
It is therefore timely to update this contribution using modern experimental input for the \mbox{$R$-ratio} and contemporary analysis techniques.
We include hadronic contributions based on the dispersive approach, 
accounting for the current tensions in different HVP evaluations (see the 2025 White Paper of the Muon $g-2$ Theory Initiative~\cite{Aliberti:2025beg,Aoyama:2012wk,Volkov:2019phy,Volkov:2024yzc,Aoyama:2024aly,Parker:2018vye,Morel:2020dww,Fan:2022eto,Czarnecki:2002nt,Gnendiger:2013pva,Ludtke:2024ase,Hoferichter:2025yih,RBC:2018dos,Giusti:2019xct,Borsanyi:2020mff,Lehner:2020crt,Wang:2022lkq,Aubin:2022hgm,Ce:2022kxy,ExtendedTwistedMass:2022jpw,RBC:2023pvn,Kuberski:2024bcj,Boccaletti:2024guq,Spiegel:2024dec,RBC:2024fic,Djukanovic:2024cmq,ExtendedTwistedMass:2024nyi,MILC:2024ryz,FermilabLatticeHPQCD:2024ppc,keshavarzi:2019abf,DiLuzio:2024sps,Kurz:2014wya,Colangelo:2015ama,Masjuan:2017tvw,Colangelo:2017fiz,Hoferichter:2018kwz,Eichmann:2019tjk,Bijnens:2019ghy,Leutgeb:2019gbz,Cappiello:2019hwh,Masjuan:2020jsf,Bijnens:2020xnl,Bijnens:2021jqo,Danilkin:2021icn,Stamen:2022uqh,Leutgeb:2022lqw,Hoferichter:2023tgp,Hoferichter:2024fsj,Estrada:2024cfy,Deineka:2024mzt,Eichmann:2024glq,Bijnens:2024jgh,Hoferichter:2024bae,Holz:2024diw,Cappiello:2025fyf,Colangelo:2014qya,Blum:2019ugy,Chao:2021tvp,Chao:2022xzg,Blum:2023vlm,Fodor:2024jyn} for more details).
Details of the hadronic contributions can be found in Section~\ref{sec::hadr}.

Updated results for the prediction of the muon lifetime and 
the Fermi constant are shown in Section~\ref{sec::GF}, and
Section~\ref{sec::concl} contains this work's conclusions.

\section{\label{sec::QED}QED contribution}

We parametrize the quantity $\Delta q$ in Eq.~(\ref{eq::tau-GF}) as 
\begin{equation}
    \Delta q =
    \sum_{i \ge 0}
    \Delta q^{(i)}
    =\sum_{i \ge 0}
    \left(\frac{\bar \alpha(m_\mu)}{\pi}\right)^i
    Y_{(i)}(\rho)\,,
\end{equation}
where the expansion parameter $\bar \alpha$ is the fine-structure constant in the $\overline{\mathrm{MS}}$ scheme at the scale $\mu = m_\mu$.
The current value for the fine structure constant in the on-shell scheme is~\cite{Mohr:2024kco}
\begin{equation}
    1/\alpha = 137.035 \, 999\, 177\, (21)   . 
\end{equation}
For the lepton masses we use the current values from Ref.~\cite{ParticleDataGroup:2024cfk},
which are given by
\begin{align}
    m_e &= 0.510\, 998\, 950\, 00\, (15) \, \mathrm{MeV}\,, \notag \\
    m_\mu &=105.658\, 3755 \, (23) \, \mathrm{MeV}\,, \notag \\
    \rho &= 483\, 633\, 169 \, (11) \times  10^{-11}.
    \label{eq:lep_masses}
\end{align}
Here the uncertainty in $m_\mu$ is important for the
determination of $G_F$, as will be discussed below.
The uncertainty in $m_e$ can be neglected.
Using the relation between $\alpha$ and $\bar \alpha(m_\mu)$ up to 
fourth order~\cite{Baikov:2012rr} we obtain 
\begin{equation*}
    1/\bar \alpha(m_\mu) =
    135.901 \, 928 \, 274\, (21).
\end{equation*}
The theoretical uncertainty due to unknown $O(\alpha^5)$
corrections in the relation between $\alpha$ and $\bar \alpha$
can be estimated as $(\alpha/\pi)^5 \log^5 (\rho)/243$ and 
is at the level of $10^{-11}$.
The uncertainty in $\bar \alpha$ can be 
neglected.

In the following, we discuss the QED corrections
to $\Delta q$ up to order $\alpha^3$.
We separate the  mass-independent and mass-dependent parts:
\begin{equation}
    \Delta q^{(i)} = \Delta q^{(i)}_0+\Delta q^{(i)}_m.
\end{equation}
At leading order, the muon decay rate is described by the function
\begin{align}
F(\rho) &=1  - 8\rho^2 - 24 \rho^4 \log \rho + 8 \rho^6 - \rho^8
\end{align}
and the finite electron mass effect at leading order is given by
\begin{equation}
    \Delta q^{(0)}_m = F(\rho)-1=-187\, 050.828 (8) \times 10^{-9}.
\end{equation}

To first order in $\alpha$ one has 
\begin{align}
    \Delta q^{(1)} &=
    \frac{\bar \alpha(m_\mu)}{ \pi}
    \Bigg[ 
    \frac{25}{8}-\frac{\pi^2}{2} 
    -\rho^2 (34 + 24 \log(\rho))
    +O(\rho^3)
    \Bigg] 
    \notag \\
    &=\frac{\bar \alpha(m_\mu)}{ \pi}
    \Bigg[\! -\!1.8098 \dots +  2.214\, 217\, 44 \, (8) \!\times\! 10^{-3} |_{m_e} \Bigg]
    \notag \\
    &= (-4\, 238 \,923.9 +5\, 186.1|_{m_e}) \times 10^{-9} 
    \notag \\
    &= 
    -4\, 233\, 737.8\times 10^{-9}\,,
\end{align}
where for the numerical evaluation of the finite electron mass effect at NLO
we used the exact analytic expression from Ref.~\cite{Nir:1989rm}.
In the final value for $\Delta q^{(1)}$ we neglect the small 
uncertainty stemming from the ratio $\rho$.
At order $\alpha^2$  the corrections are given by~\cite{vanRitbergen:1998yd,vanRitbergen:1999fi,Pak:2008qt}
\begin{align}
\Delta q^{(2)} &=
\left(\frac{\bar \alpha(m_\mu)}{\pi} \right)^2
\Bigg[
\frac{156815}{5184}
-\frac{518}{81} \pi^2
-\frac{895}{36}\zeta_3
\notag \\
&\quad 
+\pi ^2 \frac{53 \log (2)}{6}
+\frac{67 \pi ^4}{720}
-\frac{5 \pi ^2 }{4} \rho+O(\rho^2)
\Bigg] 
\notag \\
&=
\left(\frac{\bar \alpha(m_\mu)}{\pi} \right)^2
\left(6.74268\dots -0.079\, 159\, 704\, (2)|_{m_e} \right)
\notag \\ 
&= \Big( 36 \,989.8
-434.3|_{m_e}
\Big)\times 10^{-9}
\notag \\
&=36 \, 555.5\times 10^{-9}.
\end{align}
The small contributions arising at NNLO from
closed $\tau$ loops were calculated in Ref.~\cite{vanRitbergen:1998hn}
\begin{align}
    \Delta q^{(2)}_{\tau} &= 
    - 0.00058 \left(\frac{\bar \alpha(m_\mu)}{\pi} \right)^2 \notag  
    =-3.2 \times 10^{-9}.
\end{align}
At third order we have
\begin{align}
    \Delta q^{(3)}_0 &=
    \left(\frac{\bar \alpha(m_\mu)}{\pi} \right)^3
    ( 2.109799281 n_h 
    -0.01876788909 n_h^2 
    \notag \\
    &
    -7.187551125 n_l
    -6.919459635 n_l^2 
    \notag \\
    &
    -0.01288114148 n_h n_l
     -6.2 \pm 1.6 ) 
    \notag 
    \\
    &=
    \left(\frac{\bar \alpha(m_\mu)}{\pi} \right)^3
    ( -18.3 \pm 1.6 ) 
    \notag \\
    &=
   (-235 \pm 20) \times 10^{-9}\,.
    \label{eq::Dq3}
\end{align}

The fermionic corrections, marked by powers of $n_l=1$ and $n_h=1$ for closed electron and muon loops respectively,
are available with high precision from Ref.~\cite{Fael:2023tcv}.
The bosonic corrections (i.e.\ the terms independent of $n_l$ and $n_h$ in Eq.~(\ref{eq::Dq3}))
are taken from the $C_F^3$ color factor of the calculation performed in
Ref.~\cite{Fael:2020tow} for  $ b \to c l \bar \nu_l$.
Expansion terms up to $\delta^{12}$ with $\delta=1-m_c/m_b$ have been
computed in~\cite{Fael:2020tow}. 
In the limit $\delta\to 1$ we obtain $-5.31 \pm 1.33$.
In Eq.~\eqref{eq::Dq3}, we use two additional expansion terms, up to $\delta^{14}$, which have been computed since the publication of~\cite{Fael:2020tow}.

To understand the quality of the convergence of the expansion in $\delta$
we can compare, within QCD, the exact result for 
the third-order correction of $b \to u l \bar \nu_l$
from Ref.~\cite{Chen:2026jwl,Chen:2026gin} with the extrapolation obtained 
from Ref.~\cite{Fael:2020tow} using the results up to $\delta^{12}$
and setting $\delta=1$ (i.e. $m_c=0$).
For the bosonic contribution at third order $X_{3,\mathrm{bos}}$ 
we have
\begin{align}
    X_{3,\mathrm{bos}}^{\cite{Chen:2026jwl}} &=-550.014, &
    X_{3,\mathrm{bos}}^{\cite{Fael:2020tow}} &=-526 \,(38)
\end{align}
which differ by only $4.3\%$ and are compatible within uncertainty.
Note, however, that the absolute difference is dominated 
by the $C_A^2 C_F$ term, which is absent in QED.
In Ref.~\cite{Chen:2026jwl} the QCD results are not presented
for individual color factors. 
Thus, we cannot extract the color factor $C_F^3$,
which is the relevant piece for the translation from QCD to QED.

\section{\label{sec::hadr}Hadronic Contribution}

Hadronic contributions to $\tau_\mu$ enter at order $\alpha^2$ and can be calculated using the dispersion integral\footnote{A result for $Y_{(2)}^\text{had}$ could alternatively be predicted from lattice QCD.}
\begin{equation}
    Y_{(2)}^\mathrm{had}=
    \int_{m_0^2}^\infty \frac{\mathrm{d}s}{3s}R(s)K_\Gamma(s).
    \label{eqn:Y2had}
\end{equation}
Here, $m_0$ is the neutral pion mass and $R(s)$ is the hadronic \mbox{$R$-ratio}, defined as the total $e^+e^-\to\mathrm{hadrons}$ cross section $\sigma_\text{had}(s)$ normalized to the LO massless muon cross section, i.e. $R(s) = \sigma_\text{had}(s)(4\pi\alpha^2/(3s))^{-1}$. The monotonic kernel $K_\Gamma(s)$ is given by Eq.~(7) of~\cite{vanRitbergen:1998hn}. As $s\to\infty$, $K_\Gamma(s)\sim2m_\mu^2/(5s)\ln(m_\mu^2/s)$ so lower energy contributions are more heavily weighted.

The hadronic $R$-ratio used is that of KNTW~\cite{keshavarzi:2024bli,keshavarzi:2026xwi}, based on the KNT19 data compilation~\cite{keshavarzi:2018mgv,keshavarzi:2019abf}. This compilation consists of a sum of exclusive hadronic final states (channels) for $\sqrt{s}<1.937$~GeV, inclusive hadronic measurements between $1.937$~GeV and $11.199$~GeV, and perturbative QCD for $\sqrt{s}>11.199$~GeV. These are supplemented by chiral perturbation theory estimates of the thresholds of leading low energy channels and Breit-Wigner descriptions of the narrow $J/\psi$, $\psi'$ and $\Upsilon(1S-4S)$ resonances. Such additional contributions are not resolved in the experimental $e^+e^-\to\mathrm{hadrons}$ data, but are wholly subdominant.
The estimate of $Y_{(2)}^\mathrm{had}$ is dominated by the contributions from the $\pi^+\pi^-$ channel, which make up \mbox{$\sim65\%$} of the total. Uncertainties on $Y_{(2)}^\mathrm{had}$ are derived directly from the covariance matrices of the combination.

Despite the release of $>50$ new high precision datasets since KNT19, an updated $R$-ratio compilation is not presently available. The KNTW analysis framework is currently subject to software blinding and updated results cannot be obtained from it~\cite{keshavarzi:2024bli}. An important motivation for the introduction of this blinding was the persistent up to \mbox{$\sim5\sigma$} tensions between the recent \mbox{CMD-3} measurement of the $\pi^+\pi^-$ cross section ($\sigma_{\pi\pi}$)~\cite{CMD-3:2023alj,CMD-3:2023rfe} and earlier comparably precise $\sigma_{\pi\pi}$ datasets (e.g.~\cite{BaBar:2009wpw,BaBar:2012bdw,KLOE:2008fmq,KLOE:2010qei,KLOE:2012anl,KLOE-2:2017fda}). The \mbox{CMD-3} $\sigma_{\pi\pi}$ values ($\sigma_{\pi\pi}^{\text{CMD-3}}$) exceed the older data by as much as $\sim5\%$, consistently across the spectrum.  Significant efforts to try to resolve this discrepancy~\cite{Strong2020,Aliberti:2024fpq,PetitRosas:2026iuq,Budassi:2024whw,Budassi:2026lmr,CarloniCalame:2026hhy} have not yet resulted in a satisfactory explanation, though studies (including new $\sigma_{\pi\pi}$ measurements) are ongoing~\cite{CarloniCalame:2015obs,Abbiendi:2016xup,Abbiendi:2677471,BaBar2025,BelleII2025,BESIII2025,CMD32025,KLOE2025,SND2025}.

In this work, both the KNT19 combination and the \mbox{`KNT19/CMD-3'} combination~\cite{DiLuzio:2024sps} are used to compute $Y_{(2)}^\mathrm{had}$. In the latter, the $\pi^+\pi^-$ data combination for the range $0.327-1.199$~GeV (the range of the \mbox{CMD-3} measurement) is replaced with $\sigma_{\pi\pi}^{\text{CMD-3}}$. This provides a realistic upper limit on the values obtained for integrated observables such as $Y_{(2)}^\mathrm{had}$. These compilations are individually input into Eq.~\eqref{eqn:Y2had} to obtain
\begin{widetext}
\begin{align}
    Y_{(2)}^\mathrm{had}[\text{KNT19}]&=-42.18(0.05)_\text{st}(0.13)_
    \text{sy}(0.08)_\text{rad}(0.08)_\text{corr}[0.18]_\text{tot}\times10^{-3},\label{eq:KNT19_result}\\
    Y_{(2)}^\mathrm{had}[\text{KNT19/CMD-3}]&=-43.34(0.04)_\text{st}(0.25)_
    \text{sy}(0.07)_\text{rad}(0.08)_\text{corr}[0.27]_\text{tot}\times10^{-3}.\label{eq:CMD-3_result}
\end{align}
\end{widetext}
Here ``st'' and ``sy'' refer to uncertainties propagated from experimental statistics and systematics; ``rad'' refers to uncertainties on the a posteriori radiative corrections necessary in the dispersive approach; ``corr'' refers to the new KNTW uncertainty accounting for the impact of imperfectly known systematic correlations~\cite{keshavarzi:2026xwi}\footnote{Note that this uncertainty is denoted $\rho$ in~\cite{keshavarzi:2026xwi}.}; and ``tot'' denotes the total uncertainty obtained from a quadrature sum of the individual error sources (given in square brackets).

Accounting for correlation between Eqs.~\eqref{eq:KNT19_result} and~\eqref{eq:CMD-3_result}, this constitutes a $4.7\sigma$ tension. Since it will not be the leading source of uncertainty on $\tau_\mu$, it is reasonable here to take the unweighted average of these as a prediction for the central value and half of the spread as an additional uncertainty ($\pm0.58\times10^{-3}$). This conservatively accounts for all plausible results including -- as the $\pi^+\pi^-$ channel provides the dominant contribution to $Y_{(2)}^\mathrm{had}$ -- the potential impacts of other datasets missing since KNT19.
 Doing so, the result 
\begin{align}\label{eq:hadronic_contribution}
    \Delta q^{(2)}_\text{had}&=-(0.0428\pm0.0006)\left(\frac{\bar \alpha(m_\mu)}{\pi} \right)^2\notag\\
    &=-(234.6\pm3.3)\times10^{-9}
\end{align}
is obtained, the uncertainty on which is entirely dominated by the \mbox{CMD-3} tension. This result is consistent with~\cite{vanRitbergen:1998hn}, with the uncertainty reduced by a factor of \mbox{$\sim3$}. Eq.~\eqref{eq:hadronic_contribution} can be updated when the KNTW analysis is unblinded and a new combination properly inclusive of \mbox{CMD-3} can be obtained.

To estimate the size of the hadronic correction at N$^3$LO, 
we first observe that the contribution from a closed muon loop at NNLO is $\Delta q^{(2)}_\mu = -200 \times 10^{-9}$, which is comparable in magnitude to the hadronic contribution in Eq.~\eqref{eq:hadronic_contribution}.
Similar compatibility is seen for the muon loop and HVP contributions to the muon anomalous magnetic moment $a_\mu$ at orders $\alpha^2$ and $\alpha^3$~\cite{Petermann:1957hs,Kinoshita:1972erc,Cvitanovic:1974um,Levine:1973mpz,Carroll:1974bu,Carroll:1975jf,Kinoshita:1995ym,Laporta:1996mq,keshavarzi:2019abf}, with the NNLO hadronic contribution in Eq.~\eqref{eqn:Y2had} also dominated by the low-energy region.
At $O(\alpha^3)$ the contribution from closed muon loops is  $\Delta q^{(3)}_\mu = 27 \times 10^{-9}$ and, given the comparability of contributing Feynman diagrams, we expect $\Delta q_\mathrm{had}^{(3)}$ to be of the same sign and comparable size.
We therefore base an estimate of $\Delta q_\mathrm{had}^{(3)}$ on $\Delta q^{(3)}_\mu$ and assign a conservative $100\%$ relative uncertainty, i.e.
\begin{equation}
    \Delta q_\mathrm{had}^{(3)}=(27\pm27)\times10^{-9}. 
\end{equation}

\section{\label{sec::GF}Muon lifetime and Fermi constant}

Summing the uncertainties in quadrature, we obtain the final prediction
\begin{align}
    \Delta q &=(-4\, 384\, 678.5
    \pm 3.3_\mathrm{had} 
    \pm 20_\mathrm{QED} 
    \pm 27_\mathrm{h.o.\, had} )\times 10^{-9}
    \notag \\
    &=(-4\, 384\, 678 \pm 34)\times 10^{-9}.
    \label{eqn:DeltaqTot}
\end{align}
Here, the first uncertainty (``had'') comes from the NNLO hadronic contribution in Eq.~\eqref{eq:hadronic_contribution},
the second (``QED'') is induced by the uncertainty on $\Delta q^{(3)}_0$ in Eq.~\eqref{eq::Dq3}, and the third (``h.o.$\:$had'') accounts for 
unknown hadronic corrections at $O(\alpha^3)$.
It is common also to define the Fermi coupling constant
by factorizing the leading-order phase-space function $F(\rho)$, 
\begin{equation}
    \frac{1}{\tau_\mu} = 
    \frac{G_F^2m_\mu^5}{192 \pi^3}
    F(\rho) \left[ 1+\Delta q' \right].
\end{equation}
With this alternative convention, we obtain 
\begin{equation}
    \Delta q' = (-4\, 198\, 413 \pm 34 ) \times 10^{-9}.
\end{equation}

The effect of higher-order effects
that are currently not included in our calculation can be straightforwardly estimated.
At N$^3$LO, we have neglected the dependence on the electron mass. 
These terms are expected to scale as 
$\Delta q^{(3)}_m \sim(\alpha/\pi)^3 \rho \log \rho$, which gives 
$\Delta q^{(3)}_m \sim 0.3 \times 10^{-9}$.
This is about one order of magnitude smaller
than the uncertainty associated with the NNLO hadronic corrections.

Using the average value for the muon lifetime~\cite{MuLan:2012sih,ParticleDataGroup:2024cfk},
\begin{equation}
    \tau_\mu = 2.196\,9811 \, (22)  \times 10^{-6}\, \mathrm{s},
\end{equation}
together with the muon mass (see Eq.~\eqref{eq:lep_masses}), 
we can then extract the Fermi coupling constant from Eq.~\eqref{eq::tau-GF}, 
\begin{align}
    G_F &= \sqrt{\frac{192\pi^3}{m_\mu^5 \tau_\mu (1+\Delta q)}}
    \notag  \\
    &=1.166\,378\, 59  \, (59) \times 10^{-5} \, \mathrm{GeV}^{-2}. 
    \label{eq::GF}
\end{align}
We obtain the following contributions to the relative uncertainty of $G_F$:
\begin{widetext}
\begin{align}
    \frac{1}{2}\frac{\sigma_{\Delta q}}{1+\Delta q} &=
    0.017 \, \mathrm{ppm},
     &
    \frac{5}{2}\frac{\sigma_{m_\mu}}{m_\mu} &=
    0.054\, \mathrm{ppm}, &
    \frac{1}{2}\frac{\sigma_{\tau_\mu}}{\tau_\mu} &=
    0.5\, \mathrm{ppm}.
\end{align}
\end{widetext}
This comparison clearly shows that the uncertainty in the muon lifetime 
is by far the dominant source of uncertainty in $G_F$. 
The muon-mass uncertainty provides the next-largest contribution, 
while the uncertainty induced by $\Delta q$ is significantly smaller 
and remains subleading even compared with the muon-mass uncertainty.
The uncertainties associated with $\rho$ and $\bar \alpha(\mu)$
are entirely negligible.

For the total theoretical uncertainty,
the main contribution comes from the uncertainty of the third-order correction and the unknown hadronic corrections at $O(\alpha^3)$, 
with the $O(\alpha^2)$ hadronic uncertainty 
now only making up around 1.0\% of the theoretical error.
Without the improvements presented in this work,
the NNLO uncertainty induced by $\Delta q$ would be 0.15~ppm~\cite{vanRitbergen:1998yd,Pak:2008qt,MuLan:2012sih},
which is of the same order  as the uncertainty from the muon lifetime.
Our result therefore reduces the theoretical uncertainty  
by an order of magnitude.
In order to meaningfully reduce the total uncertainty of the Fermi coupling constant, 
the experimental uncertainties in the muon lifetime and muon mass must be reduced further.

\section{\label{sec::concl}Conclusions}

We present an updated value of the Fermi constant $G_F$, derived from the muon lifetime.
Our main result is given in Eqs.~(\ref{eqn:DeltaqTot}) and~(\ref{eq::GF}). 
It incorporates exact fermionic corrections up to order $\alpha^3$ (including $O(\alpha^3)$ QED corrections) and a resulting updated estimate of the remaining uncertainty.
This leads to a reduction of the theory uncertainty by a factor of nearly ten.
Furthermore, it features an updated dispersive hadronic contribution, whose uncertainty has been reduced by a factor of approximately three.
This contribution accounts for the significant tensions present in the input hadronic data, which are driven by the recent CMD-3 $\pi^+\pi^-$ cross section measurement.

Our result for $G_F$ in Eq.~(\ref{eq::GF}) is in good agreement with
the value reported in the 2026 edition of the
PDG~\cite{ParticleDataGroup:2026aaa}.
Following our improvements, the uncertainty in $G_F$ arising from higher-order corrections is a factor of three smaller than the uncertainty in the muon mass and about a factor of 30 smaller than that of the muon lifetime; this underscores the need for improved measurements of $m_\mu$ and $\tau_\mu$.
Since the uncertainty of $G_F$ is now dominated by the measurement of the muon lifetime, any improvement in this measurement leads directly to a more precise result for the Fermi constant.

\begin{acknowledgments}

AK is supported by The Royal Society (URF$\backslash$R1$\backslash$231503).
KS is supported by the European Union under the Marie Sk{\l}odowska-Curie Actions (MSCA) Grant 101204018.
MS is supported by the Deutsche Forschungsgemeinschaft (DFG, German Research Foundation) under grant 396021762 --- TRR 257 ``Particle Physics Phenomenology after the Higgs Discovery''.
TT is supported by the STFC Consolidated Grant ST/X000699/1.
AW is supported by a PGR studentship jointly funded by STFC and the Leverhulme Trust under LIP-2021-014.
\end{acknowledgments}

\bibliographystyle{apsrev4-2}
\bibliography{bib}
\end{document}